\documentclass[journal,double]{IEEEtran}

\usepackage[utf8]{inputenc}
\usepackage{graphicx}
\usepackage{color}
\usepackage{xcolor,colortbl}
\usepackage{amsmath}
\usepackage{graphicx}
\usepackage{multicol}
\usepackage{array}
\usepackage{mathrsfs}
\usepackage{multirow}
\usepackage{xcolor}
\usepackage{lipsum}
\usepackage{titlesec}
\usepackage{mathtools}
\usepackage{comment}

\titlespacing\section{0pt}{2pt plus 2pt minus 0pt}{2pt plus 2pt minus 2pt}
\titlespacing\subsection{0pt}{2pt plus 2pt minus 2pt}{2pt plus 2pt minus 2pt}
\titlespacing\subsubsection{0pt}{2pt plus 2pt minus 2pt}{2pt plus 2pt minus 2pt}

\newcommand{\ignore}[1]{{}}

\title{\LARGE{Deep Decarbonization of Multi-Energy Systems: A Carbon-Oriented Framework with Cross Disciplinary Technologies}}

\author{Jian Shi, Dan Wang, Chenye Wu, and Zhu Han\vspace{-9mm}
}
\begin{document}

\maketitle

\begin{abstract}
The retirement of unabated coal power plants, the plummeting cost of renewable energy technologies, along with more aggressive public policies and regulatory reforms, are occurring at an unprecedented speed to decarbonize the power and energy systems towards the 2030 and 2050 climate goals. This article aims to establish a carbon-oriented framework to examine the role carbon emission is playing within a power grid that is rapidly transitioning to an integrated multi-energy system. We divide the carbon flows in the multi-energy systems into three stages: carbon allowances initialization/allocation, exchanging/pricing, and circulation. Then, we discuss how cross-disciplinary technologies, such as game theory, optimization, and machine learning can facilitate the modeling and analysis of the proposed framework.

%decarbonize the power and energy systems towards
\end{abstract}

\begin{IEEEkeywords}
Carbon footprint, Energy transition, Deep decarbonization, Multi-energy systems, Climate change
\end{IEEEkeywords}

\section{Introduction}

Electric power systems are the backbone of the global energy system and account for 25\% of worldwide greenhouse gas (GHG) emissions. Driven by the urgent need to combat climate change \cite{IPCC2021}, the electric power industry has started to shift its focus towards the near-term urgency of deploying \textit{deep decarbonization} to meet the immediate emission reduction goal by 2030 \cite{DeepDecarbon} while fostering a pathway for a much more profound and deeper zero-emission transformation for long-term climate safety by 2050. The electricity grid is rapidly transitioning to an integrated multi-energy system with the electricity system being the core, renewable energy as the primary energy sources, and supported by storage, hydro, bioenergy, natural gas, responsive demand, and electric vehicles \cite{MAS}. A low-carbon electric power system will also enable the clean electrification of transportation, buildings, and industry sectors to support full economy-wide decarbonization.

%Electric power systems are the backbone of the global energy system and account for 25\% of worldwide greenhouse gas (GHG) emissions. While most existing studies consider 2050 as the focus of the deep decarbonization of power and energy systems, it is recently estimated by the United Nations Intergovernmental Panel on Climate Change (IPCC) that the global carbon emissions must be reduced by 50\% by 2030 to limit warming to 1.5°C and avoid catastrophic climate consequences \cite{IPCC2021}. Attention has since been shifting to the near-term urgency of deploying \textit{deep decarbonization} to meet the immediate emission reduction goal by 2030 \cite{DeepDecarbon} while fostering a pathway for a much more profound and deeper zero-emission transformation of the electricity grid for long-term climate safety. Driven by the urgent need to meet the 2030 emissions target, the electricity grid is rapidly transitioning to an integrated multi-energy system with the electricity system being the core, renewable energy as the primary energy sources, and supported by storage, hydro, bioenergy, natural gas, responsive demand, and electric vehicles \cite{MAS}. A low-carbon energy system will also enable the clean electrification of transportation, buildings, and industry sectors to support full economy-wide decarbonization.

However, there are many challenges to achieve these ambitious goals. The existing electric power system is characterized by long-term assets, well-established regulatory structures, and rigid operational conventions. Moving forward, the reinvention of the traditional electricity system to integrate the complex dynamics of multi-energy supply and carriers has brought fundamental changes for the power community who used to only concern the “electrical” aspect of the system. The planning, operation, and modeling of the future multi-energy system necessitate stronger and more holistic collaboration across multiple energy sectors to maintain the desired operational characteristics and societal responsibilities, with enhanced energy efficiency, flexibility, and affordability. Meanwhile, with decarbonization as the key priority, a carbon-centered systematic framework also becomes necessary to pave the pathway of the low/zero-carbon transition of the multi-energy system and develop fundamentally new methods, as well as modifications and supplements to existing conventions, to meet the 2030 climate goal.

\begin{figure}[t!]
    \centering
    \setlength{\belowcaptionskip}{-3mm}
    \includegraphics[width=1\columnwidth]{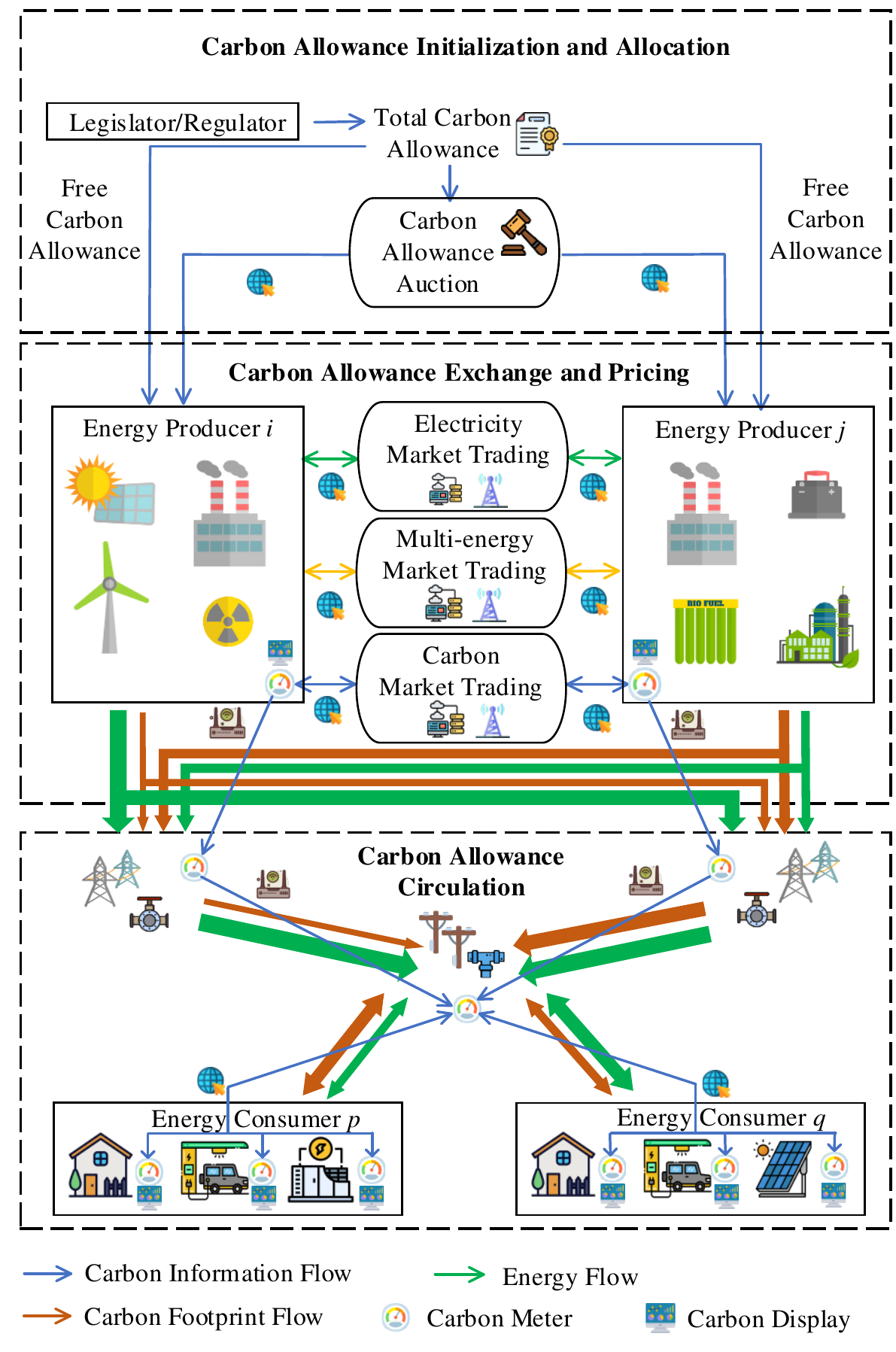}\vspace{-3mm}
    \caption{Carbon flows through the multi-energy system}
    \label{fig1}
\end{figure}

%To link MES and emission policies, information needs to be obtained and utilized so as to optimize and learn the best choices for carbon neutralization, while little literature has this cross disciplinary viewpoint. 

%Deep decarbonization in the complex multi-energy system urgently requires an organizing structure to provide a holistic view on the agents, components, resources, and goals of the respective parties. 
This article aims to establish a carbon-oriented framework to facilitate the multi-energy system’s effective decarbonization and deliver reliable, secure, clean, and cost-effective energy services. In particular, motivated by the tremendous role decarbonization will be playing in the future energy system, we take a different route than the traditional literature, which primarily scoped from the electricity and energy perspective concentrating on one piece of technological solutions. We focus on developing a carbon perspective to analyze and exploit the synergy between the ``life cycle" of carbon emissions in the energy value chain. Specifically, we divide the carbon flows through the multi-energy system into three stages (see Fig. \ref{fig1}): \textbf{carbon allowance initiation and allocation}, \textbf{carbon allowance exchange and pricing}, and \textbf{carbon allowance circulation}, based on which interdisciplinary researches, e.g., optimization, game theory, and machine learning, can be performed to collectively accelerate the carbon neutral transition of the energy sector as discussed in Section \ref{sec:technologies}.

%, comprehensive and disruptive policy package and other measures, harvest the low-hanging fruits while picking the high-hanging fruits as well
%a decarbonized grid

\section{Carbon Allowance Initiation and Allocation}
\label{sec:carbon_allowance_initiation}

To establish a cap-and-trade carbon-emission trading (CET) system, the first step for a regulator is to initialize the emission cap based on its emission reduction target as well as other policy-level assessment, such as technology readiness, mitigation potential, and costs. Then, the capped amount of carbon allowances, as an asset, has to be assigned to the market participants. The choice of this allocation directly affects how the energy producers participating in the CET react to the carbon price signal and then determine their optimal operational and investment strategies. Therefore, the selection of the correct allocation strategy is the key to establish a well-functioned and competitive market,  preserve the appropriate incentives for emission mitigation, and ensure economic efficiency. Currently, there existing two general strategies to allocate the carbon allowance: %free allocation through grandparenting and allocation through Auctioning.

%Additionally, the allowance allocation also enables effective management of the CET for the policy maker to stabilize carbon pricing, prevent undesirable market impacts, and reward early reduction actions.

\textbf{\em Strategy 1. Free allocation through grandparenting}: In this approach, carbon allowances are allocated to energy producers completely free based on historical benchmarks, such as emission intensity levels and power/energy generation, as well as an \textit{assistance rate}. The assistance rate is used as a policy-level tool to scale  the level of emissions that receive a free allocation. While the grandparenting approach is straightforward and easy to implement, it draws constant criticism since grandfathering all carbon allowances creates a large windfall profits for energy producers. This is because the marginal cost resulted from the CET system would be ultimately passed to the consumers in the form of higher electricity prices. Therefore, giving away the allowances would hardly impose any additional cost to the energy producers, resulting in increased profits and asset values. Given its drawbacks, free allocation through grandparenting is considered a transitional approach. For instance, it has only been used in the EU ETS in Phase 1 (2005-2007) where almost all the allowances were given out for free and Phase 2 (2008-2012) where free allocation fell to around 90\% of the total allowances. In Phase 3 (2013-2020) of the EU ETS, auctioning became the default method for allocating allowances \cite{ETS_web}\cite{EU_ETS}.

%For instance, the assistance rate can be set higher (e.g., 90\%) for highly emission-intensive activities and lower (e.g., 40\%) for less emission-intensive activities. 

\textbf{\em Strategy 2. Allocation through auctioning}: In the auctioning approach, all the market-participating energy producers have to purchase carbon allowances as \textit{buyers} from an auctioning platform (e.g., European Energy Exchange (EEX) used by EU ETS). On the other hand, the total amount of allowances that can be auctioned is distributed to multiple stakeholders who own auction rights (e.g., Member States of EU in the EU ETS). These stakeholders participate in the auction process on the same auctioning platform as \textit{sellers}.

%In practice, a sensible mix of the above mentioned allocation strategies would be necessary to facilitate efficient near-term and foster long-term technological innovations towards the goal of deep decarbonization. For instance, in Phase 3 of the EU ETS, roughly 43\% of the total cap was given to the industry sector for free. However, the electric power sector was subject to 100\% auctioning. Moreover, 5\% of the allowances were set aside in the ``new entrants reserve" to be assigned for free to encourage the deployment of innovative, renewable energy technologies and carbon capture/storage \cite{EU_ETS}.

\section{Carbon Allowance Exchange and Pricing}
\label{sec:carbon_allowance_exchange}
The cross-energy-sector nature of the multi-energy system has made it evident that an energy producer can benefit both financially and functionally by participating in multiple energy markets as well as the carbon emission market. More specifically, strategic energy producers that own electric and other energy supplies and demands can participate jointly in multiple markets (e.g., electricity market and natural gas market) by submitting offers and bids to maximize their profits. Energy producers can also exercise market power and trade their carbon allowances obtained through free allocation or auctioning in the CET system. In this way, markets across different energy sectors and CET become coupled. This process can be described as follows:

\textbf{\em For each energy producer}, strategic energy producer $i$ seeks to maximize its own profits from participating in a number of markets, including electricity, natural gas, hydrogen, and carbon markets simultaneously, as well as cost resulted from cross energy sector operations. Each energy producer is also subjected to a set of operational constraints, including sector-specific operational constraints, cross-sector operational constraints, and carbon-related constraints, e.g., the carbon balance between carbon allowance and carbon consumption, as well as other low-carbon operational requirements.

\textbf{\em For each market operator}, take the electricity market as an example, an electricity market operator seeks to maximize the energy social welfare, including the combined utility of the power demand minus the total cost of power production. Each market is also characterized by its market-specific constraints. For an electricity market, such constraints include power balance, operation limits and bounds for market participants, capacity of transmission lines, and voltage angle for an electricity market. Similarly, for the carbon market, the carbon market operators need to enforce a set of carbon-related constraints, such as ensuring the carbon balance within the market, carbon price having to be positive, and the total carbon consumption being capped, and etc.

\section{Carbon Allowance Circulation}

We consider that the carbon flow in a multi-energy system can be modeled in two layers: 1) a \textit{carbon footprint flow} (CFF) that describes how the carbon footprints circulate with the physical energy flow in the network, potentially across multiple energy sectors; and 2) a \textit{carbon information flow} (CIF) that focuses on capturing the information exchange based on the carbon monitoring, recording, and visualization devices and mechanisms within the multi-energy system. 

\textbf{\em Carbon footprint flow}:
A modeling strategy has been proposed \cite{Carbonflow} to model and trace carbon footprints moving from various types of energy sources to energy demands in the form of an ``intertwined" carbon-energy flow. CFF includes the following essential components: 1) A carbon emission flow rate that measures the volume of carbon footprints flowing through a node or a branch; 2) A carbon emission intensity associated with energy production; 3) A carbon emission intensity associated with energy flowing along a branch (e.g., transmission line, gas pipeline, or hydrogen pipeline); 4) A carbon emission intensity associated with energy injection; 5) A carbon emission intensity associated with energy input/output of energy converters (when one energy form is converted to another). 

Based on the components mentioned above, the carbon footprint circulation in the multi-energy network can be quantified. Once the physical energy flow is determined, the CFF model can be modeled within each type of energy network, as well as across energy networks that are coupled. The CFF model can be useful for a variety of applications. For instance, it provides insights to the customers on the carbon footprints of their electricity consumption which may stimulate demand-side emission mitigation. CFF can also be used by policymakers to better allocate carbon allowances as well as the trading market design.

\textbf{\em Carbon information flow}:
%The multi-energy power grid consists of a large number of autonomous systems, including renewable energy systems as wind farm, hydro plant, PV farm, etc., fossil fuel energy systems as thermal plant, gas plant, etc, energy generation and distribution systems as power plants, heat networks, and gas networks, as well as energy consumer systems.
The integration of multiple energy forms in the multi-energy system provides flexibility and immense potentials in energy and carbon saving, yet the massive amount of information devices embodied in different energy sectors also substantially increases the complexity of keeping track of the dynamic carbon flows in a rapidly evolving energy system. Therefore, in addition to the CFF, it is also necessary to develop a CIF model which records the information on the carbon circulation in fine-granularity.

To materialize CIF, an overarching information infrastructure should be developed based on the integration of the information sub-systems among different energy systems to collect, store, distribute and process carbon data for providing information, knowledge and decision-making. Such a system may consist of: 1) the sensing infrastructure on carbon metering, 2) the architecture of the information systems so that logic layers of the information systems can be formed to hide the heterogeneity of other layers and modules, 3) the protocols for the information to be easily exchanged by different energy information systems, and 4) the computing framework for distributed and collaborative decision making. 

\section{Links to Enabling Technologies}\label{sec:technologies}

Based on the three problem identified above, we illustrate the links to the cross disciplinary techniques such as computer science, big data, machine learning, game theory, statistics, and operational research societies.

%\subsection{Game: MPEC/EPEC and Auction}

\subsection{Distributed Game Theoretical Approaches}

Since not all the players participating in the deep decarbonization of the multi-energy system belong to the same authority, game theory is a natural and powerful framework with a set of mathematical tools to study the complex interactions among interdependent rational players. 
Various  concepts from game theory are adequate for the formulated problems in previous sections, e.g., equilibrium solutions that are desirable in competition scenarios since they allow designs that are robust to the deviations of any player.
Numerous solutions are available in the literature for different scenarios, such as noncooperative games, cooperative games, Bayesian games, differential games/mean field games, stochastic games, evolutionary games, matching games, contract theory, game with bounded rationality, learning in games, auction theory and mechanism design. %\cite{game_book1, game_book2}.

\begin{figure}[t!]
    \centering
    \setlength{\belowcaptionskip}{-3mm}
    \includegraphics[width=1\columnwidth]{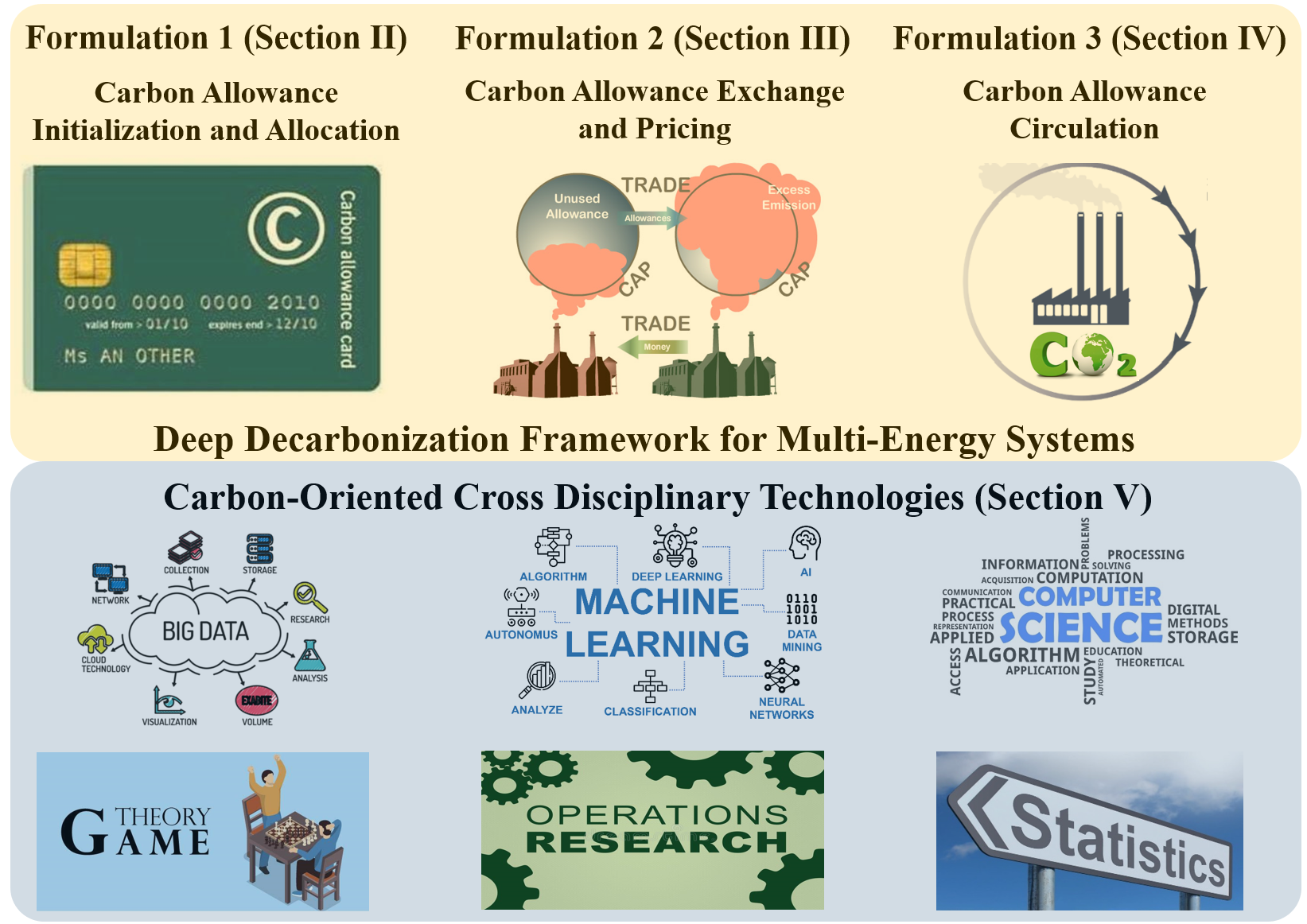}\vspace{-3mm}
    \caption{Cross-disciplinary technologies to empower the decision-making process involved in the deep decarbonization of multi-energy systems}
    \label{fig2}
\end{figure}

\subsection{Parallel and Distributed Computing for Big Data Analysis} 
The challenges in managing and analyzing ``big data" from carbon neutralization can require fundamentally new technologies in order to handle the size, complexity, or rate of availability of these data. 
Rather than developing super machines to speed up massive data processing, the choice nowadays is to use a massive number of physical machines (e.g. MapReduce and Spark) to process data in parallel. 
A large family of algorithms have been developed for big data optimization, such as Alternating Direction Methods of Multipliers (ADMM), Block Coordinate Descent Methods (BCD), sparse optimization, sublinear algorithm, and tensor optimization. %\cite{big_data_book}

\subsection{Artificial Intelligence through Federated Learning/Analysis}

Carbon information systems provide ``big data", which bring about both opportunities and challenges. For example, strategic producers need to forecast the profit obtained from electric, natural gas, and hydrogen markets, %i.e., $f_e^{(i)}({\boldsymbol{x}}_e^{(i)})$, $f_{ng}^{(i)}({\boldsymbol{x}}_{ng}^{(i)})$, and $f_h^{(i)}({\boldsymbol{x}}_h^{(i)})$. 
Machine learning models can be developed with fine-grained data in history. Model accuracy can be improved with various techniques, e.g., by leveraging domain knowledge embedded through metadata \cite{Metadata}.

There is also a critical challenge: in a multi-energy system, data are collected by the respective energy systems from diverse owners, and thus there are privacy concerns on data sharing. A viable solution approach is federated learning and analysis \cite{Federated}. Rather than sending all data to a centralized server for model training, the global model is distributed to local parties where they use their data to update the model (more specifically, the gradients) locally and return the model to the server for aggregation. This process is carried out iteratively until model convergence. No raw data on the profit of electric, natural gas and hydrogen markets of each individual energy systems is exposed.

%Formally, let min $f_i(x_i, \omega_i; y_i)$ be an update in a local party $i$. The global weight to be $\omega^s = \frac{\sum D_i \omega_i^{\tau}}{D}$. The overall objective is min $J(\omega^s) = \frac{1}{N}\sum f_i(\omega_i^{\tau})$. -- do we really need notations?

%\subsection{Uncertainty Management}

%\input{Chenye}

\bibliographystyle{IEEEtran}
\bibliography{reference}

\end{document}